\documentclass[9pt,conference]{IEEEtran}

\usepackage[preprint]{waspaa25}
\usepackage{cite}
\usepackage{amsmath,amssymb,amsfonts}
\usepackage{algorithmic}
\usepackage{graphicx}
\usepackage{textcomp}
\usepackage{xcolor}

\newcommand*{\mv}[1]{\mathbf{#1}}
\newcommand*{\mmat}[1]{\mathbf{#1}}

\newcommand*{\herm}{^{\mathsf{H}}}

\title{Accelerating Audio Research with Robotic Dummy Heads}

\name{Austin Lu$^{1}$,
      Kanad Sarkar$^{1}$,
      Yongjie Zhuang$^{2,5}$,
      Leo Lin$^{1}$,
      Ryan M. Corey$^{3,4}$,
      Andrew C. Singer$^2$
      \thanks{This research was supported in part by the Foxconn Interconnect Technology sponsored Center for Networked Intelligent Components and Environments (C-NICE) at the University of Illinois at Urbana-Champaign and by a Science Team Seed Grant from the Discovery Partners Institute. This work does not relate to the author’s position at Amazon.}
      }
\address{$^{1}$Electrical and Computer Engineering, University of Illinois Urbana-Champaign, Illinois, USA \\
$^{2}$Electrical and Computer Engineering, Stony Brook University, New York, USA \\
$^{3}$Electrical and Computer Engineering, University of Illinois Chicago, Illinois, USA \\
$^4$Discovery Partners Institute, Illinois, USA \;
$^{5}$Amazon Web Services, Seattle, USA
}
\begin{document}
\maketitle

\begin{abstract}
This work introduces a robotic dummy head that fuses the acoustic realism of conventional audiological mannequins with the mobility of robots. The proposed device is capable of moving, talking, and listening as people do, and can be used to automate spatially-stationary audio experiments, thus accelerating the pace of audio research. Critically, the device may also be used as a moving sound source in dynamic experiments, due to its quiet motor. This feature differentiates our work from previous robotic acoustic research platforms. Validation that the robot enables high quality audio data collection is provided through various experiments and acoustic measurements. These experiments also demonstrate how the robot might be used to study adaptive binaural beamforming. Design files are provided as open-source to stimulate novel audio research.
\end{abstract}

\begin{IEEEkeywords}
Audio processing, microphone arrays, robotics
\end{IEEEkeywords}

\section{Introduction}
To evaluate a speech enhancement or source separation system, researchers commonly perform experiments with loudspeakers in place of talking people. As the loudspeakers do not move, a researcher can separately record the target source image and noise/interference sounds, after which the audio can be scaled and summed to simulate various SNR conditions without the need for additional recordings. Direct access to the target and noise signals is also required to calculate many fundamental objective performance metrics \cite{CHIME1, Loizou2007SE, vincent2006BSS, beamspace}. 

However, using a loudspeaker in place of a talking person is not acoustically realistic, as there is significant mismatch in acoustic directivity of the two. Some acoustic mannequins such as the KEMAR, which provide a realistic head-related transfer function (HRTF) for recording binaural audio and testing audio devices \cite{kemar}, can be equipped with mouth simulators with humanlike directivity. Although such a mannequin can
be used in place of loudspeaker, this is rarely feasible due to the high cost, size, and weight.

Regardless of the use of human subjects, loudspeakers, and/or mannequins, it is tedious and time consuming to record audio experiments. In the current age of data-driven audio processing \cite{richard2023audio}, it would be invaluable to reduce the difficulty of data collection by automatically arranging and recording sources and microphones. Various researchers have proposed to use robots for this task, with the aim of increasing the pace of audio research \cite{leroux2015micbot, kujawski2024miracle, lu2022mechatronic, horaud_nao,deleforge2012cocktail,
deleforge2015acoustic, thesis}.

Another benefit of robots is their potential to emulate realistic scenarios with motion in a precise and repeatable manner. This is impossible for human subjects, regardless of whether the subjects move according to a script \cite{fejgin2021comparison} or naturally \cite{chime5, corey2022adaptive}, as they would have to replicate various subconscious movements across the target and mixture recordings. While robots have been used in high quality spatially-dynamic audio experiments \cite{lollmann2018locata, evers2020locata}, they have not yet been used in \textit{repeatable} dynamic experiments.
We believe that this is because of the significant challenge posed by audible motor noise \cite{schmidt2016ego}. Such noise is problematic due to its intensity, harmonic structure, and time-varying motion-dependent modulation \cite{dale1987gear}. While denoising methods can alleviate this issue, such processing can easily introduce distortion or artifacts, reducing the realism of the experiment \cite{vincent2018audio}.

Therefore, repeatable dynamic audio experiments must be simulated rather than physically recorded, for instance by convolving clean speech with a spatially dense set of room impulse responses. The room impulse responses may themselves be simulated or measured \cite{CHIME2}. In many cases, researchers will place greater emphasis on realism, and opt for real recordings with human subjects while accepting the inherent non-repeatability \cite{vincent2012signal}. Yet, if motor noise can be adequately suppressed without aggressive post-processing, repeatable audio experiments with motion could be performed.

In this work, we propose a new research tool: the robotic acoustic dummy head\footnote{https://github.com/Audio-Illinois/robot-acoustic-head}. To be of use in various audio experiments, including spatially-dynamic scenarios, this device moves precisely and quietly while exhibiting a lifelike HRTF. Acoustic measurements are provided to validate that these criteria are met. To showcase the utility of the proposed robot, preliminary results from a spatially dynamic binaural beamforming experiment are presented as well. It is the authors' hope that this device will stimulate the development of other such audio-specialized robots and facilitate novel developments in audio processing research.

\begin{figure}[t]
\begin{minipage}[b]{.32\columnwidth}
  \centering
  \centerline{\includegraphics[height=3.5cm]{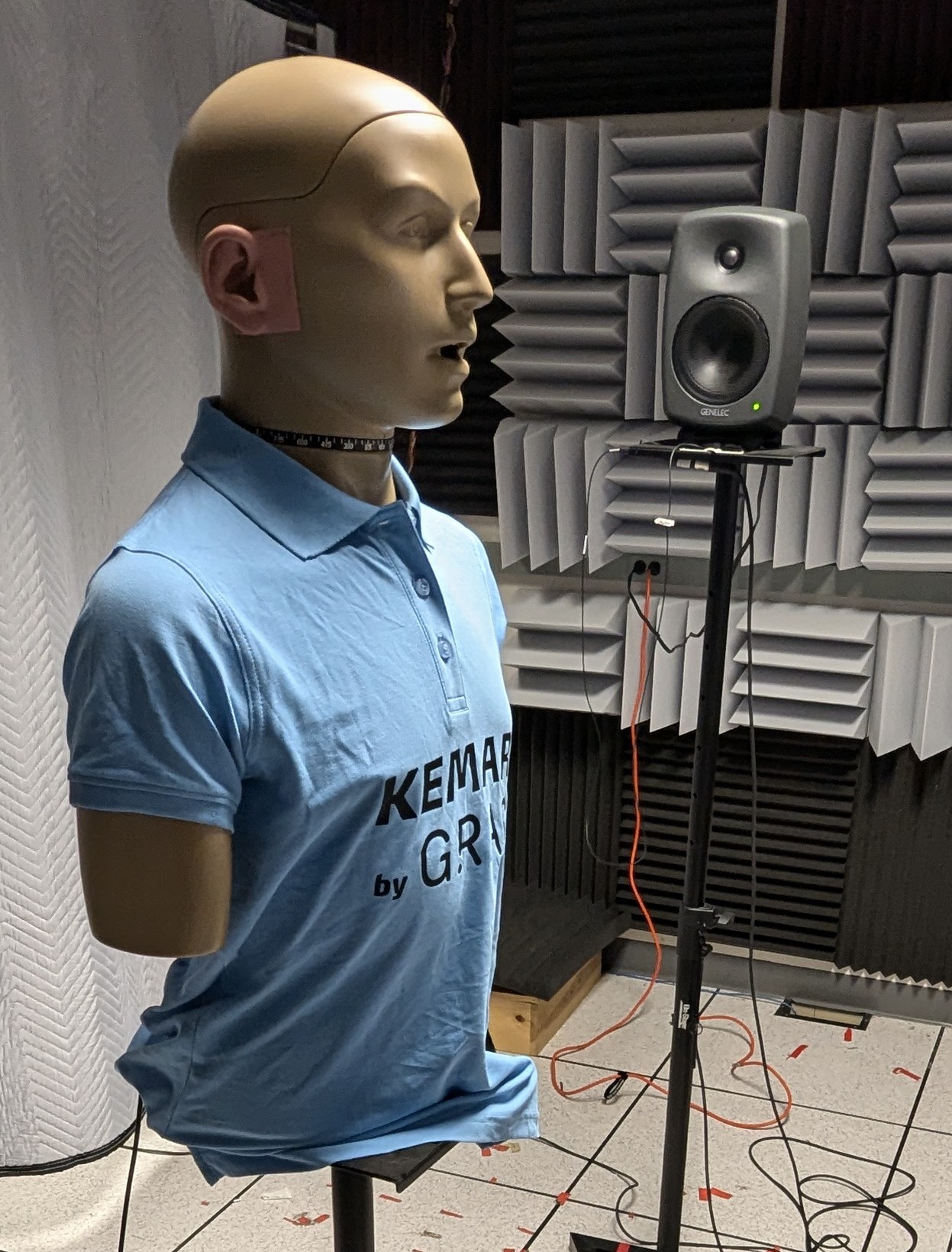}}
  \centerline{(a) KEMAR}\medskip
\end{minipage}
\begin{minipage}[b]{.32\columnwidth}
  \centering
  \centerline{\includegraphics[trim={0 0 0 9cm},clip,height=3.5cm]{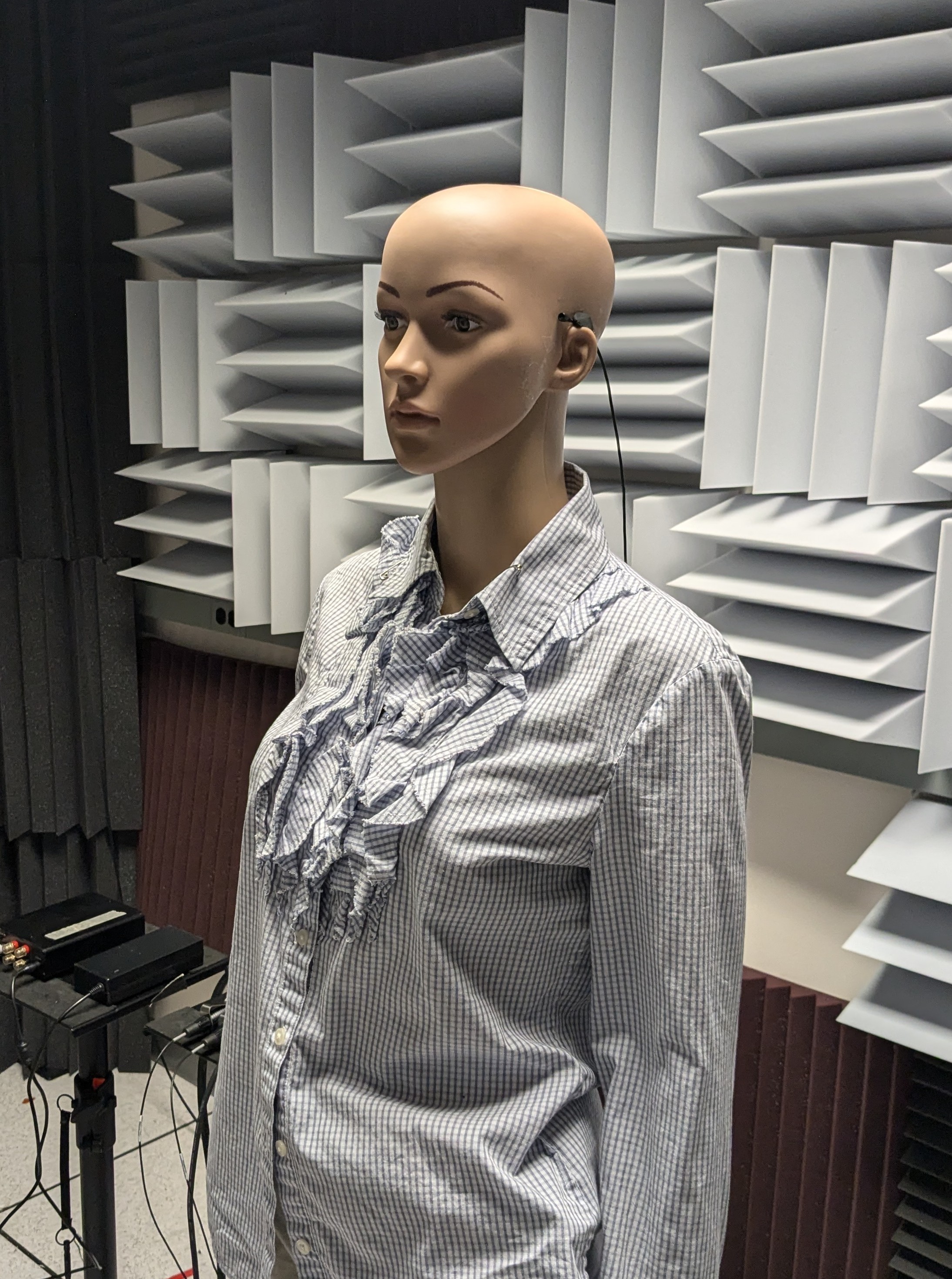}}
  \centerline{(b) Store Mannequin}\medskip
\end{minipage}
\begin{minipage}[b]{.32\columnwidth}
  \centering
\centerline{\includegraphics[trim={0cm 0 0cm 0cm},clip,height=3.5cm]{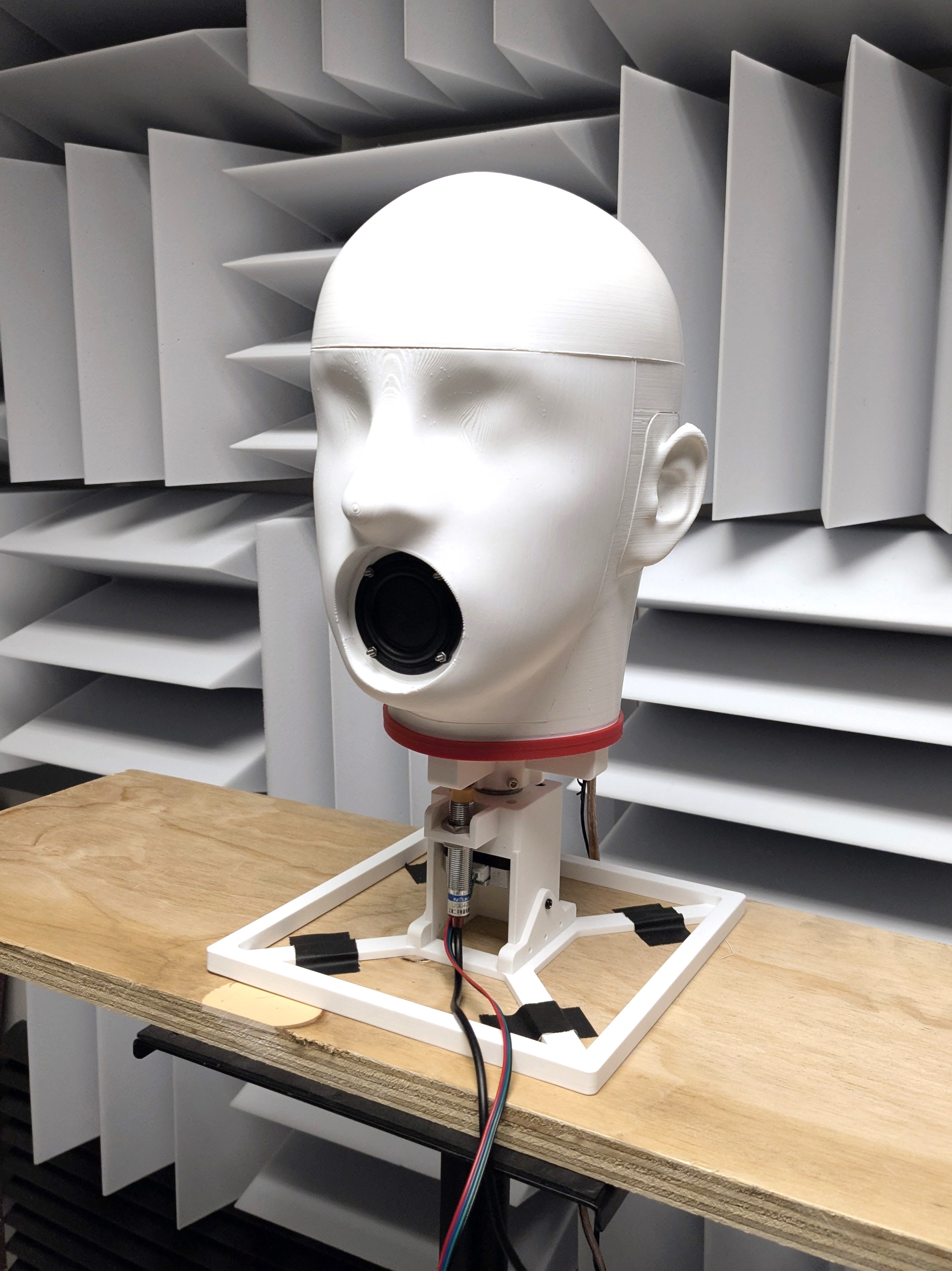}}
  \centerline{(c) Printed Head}\medskip
\end{minipage}
\caption{Mannequins such as (a) and (b) are immobile, whereas the 3D printed dummy head is lightweight and compatible with a quiet motorized platform.}
\label{fig:subjects}
\end{figure}

\section{Robotic Acoustic Dummy Head}
\label{sec:robohead}

The proposed research tool consists of two parts: the acoustically realistic dummy head and the acoustically unobtrusive turntable.

\subsection{3D printed acoustic dummy head}
\label{ssec:printhead}
An acoustic mannequin such as a KEMAR is often used in audiology to provide a standardized HRTF, which is sometimes necessary in spatial audio or for clinical purposes.
Fortunately, the level of acoustic realism achieved by these high-end calibrated research tools is not always necessary to study audio signal processing systems such as binaural beamformers or source separation algorithms.

It was shown in \cite{corey2019acoustic} that a retail mannequin provides reasonable acoustic shadowing for the study of audio-capable wearable devices across the head and body. As an improvement, \cite{yue20233d} proposed a 3D-printed dummy head. In this work, we confirm that such a design can offer many of the features of the KEMAR at a fraction of the cost. Importantly, the printed head is also smaller and has lower mass, and therefore can be maneuvered more easily than the full-body mannequins shown in \cref{fig:subjects}.

\begin{figure}[ht]
    \centering
    \centerline{\includegraphics[trim={0 0.3cm 0 0.4cm},clip,width=\columnwidth]{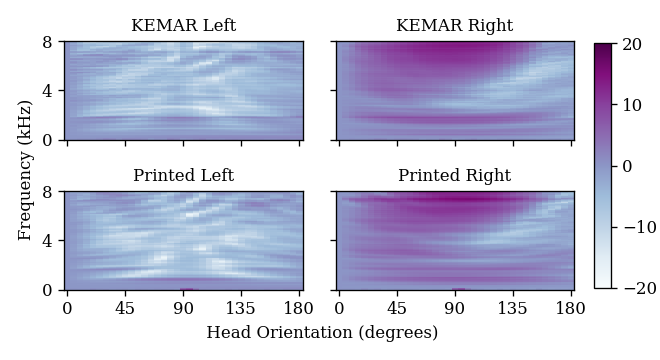}}
    \caption{
    HRTF magnitude in dB relative to 0$^\circ$ azimuth for a source 1.0 m away. The HRTFs of the KEMAR and printed head are alike at speech-relevant frequencies, indicating good acoustic realism. Note that HRTFs vary significantly across real human subjects.
    }
\label{fig:hrtf}
\end{figure}

Two omnidirectional Countryman B3 Lavalier microphones, placed in the faux ear canals of the printed head, are used to obtain binaural audio. With this approach, the HRTF of the printed head is measured in an acoustically-treated recording space at a resolution of 5$^\circ$. This process was repeated with a calibrated GRAS 45BC KEMAR to yield \cref{fig:hrtf}, which shows that the respective HRTFs are qualitatively similar, indicating a high level of acoustic realism of the printed head.

\begin{figure}[ht]
    \centering
    \centerline{\includegraphics[trim={0 0.4cm 0 0},clip,width=0.88\columnwidth]{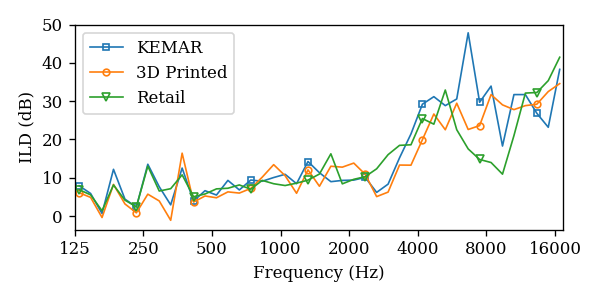}}
    \caption{ILDs measured at 90$^\circ$ azimuth for the KEMAR, printed head, and retail mannequin. Comparison of the ILDs indicates that the printed head is acoustically realistic, based on its similarity to a KEMAR.}
\label{fig:ild}
\end{figure}

For clarity, the interaural cues corresponding to a head pose of $90^\circ$ are also compared. Shown in \cref{fig:ild} are the interaural level differences (ILD) for the printed head, KEMAR, and a retail mannequin. Similarity of the printed head and KEMAR ILDs reinforce that the printed head is acoustically realistic.
The interaural time difference (ITD) between the KEMAR and printed head also shows good agreement, with a negligible difference of 62.5 $\mu$s at 90$^\circ$, and a root mean-squared error (MSE) of 67.9 $\mu$s across head orientations of $\{ 0, 5, \dots, 180 \}^\circ$.

\begin{figure}[ht]
    \centering
    \centerline{\includegraphics[trim={0 0.45cm 0 0.3cm},clip,width=0.75\columnwidth]{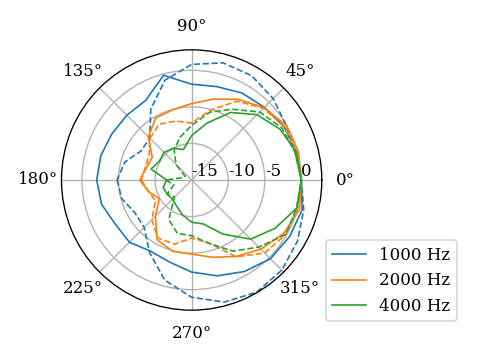}}
    \caption{
    Radiation patterns of the printed head (solid) and KEMAR (dashed), in dB relative to 0 $^\circ$ azimuth, for three octave bands.
    }
    \label{fig:dir}
\end{figure}

Unlike many dummy heads, the proposed design is also equipped with a loudspeaker to simulate the speech of a human talker. The mouth simulator design is related to that of the bespoke dummy head used in \cite{corey2021comparison}. In \cref{fig:dir}, the radiation pattern of the printed head is compared to that of a KEMAR 45BC speech simulator. Generally, high-end mouth simulators are specialized for humanlike acoustics, and thus have frequency-dependent directivity \cite{halkosaari2005directivity}, but are heavy and costly. In contrast, studio monitors are relatively maneuverable and low-cost, yet are designed for flat spectra and linearity, not lifelike radiation. The proposed design provides the benefits of both while addressing their respective drawbacks. Note that no electroacoustic surrogate for a human talker is able to accurately model the speech-dependent time-varying directivity of real people \cite{monson2012horizontal}.

To ensure that the printed head can be: 1) readily equipped with internal microphones and loudspeakers, 2) moved during recording while remaining stable, and 3) fabricated on standard $200\times 200$ mm 3D printers, the proposed design is modular with interlocking parts.

\subsection{Quiet motors, quiet robots}
\label{ssec:quiet}
A drawback of audio experiments where loudspeakers and/or mannequins are used in place of talking people is that these sound sources are immobile whereas people move constantly, for example by turning to face different conversation partners or looking around a room to localize a sound.
The proposed device performs rotation in the horizontal plane. This resembles the capabilities of the turntables and rotating platforms that are often used to automate measurements of loudspeaker and microphone directivity \cite{farina2000simultaneous}, or for collecting datasets of room impulse responses at scale \cite{psu24}.
However, these conventional devices produce significant noise when moving, so that recordings must occur \textit{between} movements. 
In general, robots are difficult to use in audio and acoustics research because both their motors and fans introduce noise \cite{ince2010hybrid}.

By using a small, low-power motor, the proposed device can rely on passive cooling and thus avoid fan noise. To address motor noise, a direct-drive (gear-free) stepper motor is used. While servomotors are generally favored for their high torque-to-weight ratio, they owe this power density to their gearboxes that are known to add severe vibrations and thus audible noise \cite{e22111306}. Stepper motors, despite a lack of gears and relatively quiet operation, can still cause problematic vibration \cite{archibald2008efficient}, so the described system also uses a specialized motor control algorithm, detailed in the design files.
An added benefit of using a stepper motor is that the nominal motor positions can be used directly, without a closed-loop control scheme. Such an approach has precedent in other motorized acoustic workbenches \cite{6784276}.

\begin{figure}[!h]
    \centering
    \centerline{\includegraphics[trim={0 0.3cm 0 0},clip,width=0.88\columnwidth]{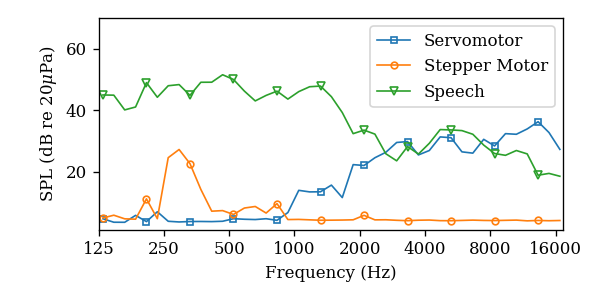}}
    \caption{ Servomotor noise is broadband and has significantly higher intensity than the stepper motor when both driven at 1 rev/s. The author's speech, at a conversational level and recorded from 1.3 m, is included for reference.}
    \label{fig:servo}
\end{figure}

That a stepper motor is significantly quieter than a servomotor is confirmed in \cref{fig:servo}. In these measurements, the effect of background noise (significant below 500 Hz) is removed by spectral subtraction. A microphone placed 1.0 m away is used to record from the acoustic far-field. The captured audio is scaled according to a calibration measurement referenced to a SPL meter with a minimum reading of 30 dBA re 20 $\mu$Pa to allow for spectral estimation.

\begin{figure}[ht]
    \centering
    \centerline{\includegraphics[trim={0 0.3cm 0 0},clip,width=0.88\columnwidth]{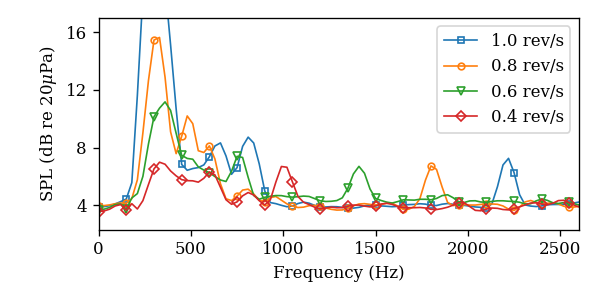}}
    \caption{ The harmonics increase in frequency proportional to the driving speed. Higher driving speeds also cause the motor to produce higher intensity sound. Yet, even for faster speeds, the quiet actuator does not exceed the level of a whisper \cite{whisperloudness}. }
    \label{fig:speed}
\end{figure}

A closer inspection of the motor noise measurements in \cref{fig:speed}, reveals that the motor produces modest harmonic noise at intensities proportional to the speed, corroborating the observations in \cite{dale1987gear}. Based on this measurement, motor speed is restricted below 0.4 revolutions per second (rev/s) where the noise is relatively spectrally white. These speeds align well with natural head movements.

\begin{figure}[ht]
    \centering
    \centerline{\includegraphics[trim={0 0.3cm 0 0},clip,width=0.88\columnwidth]{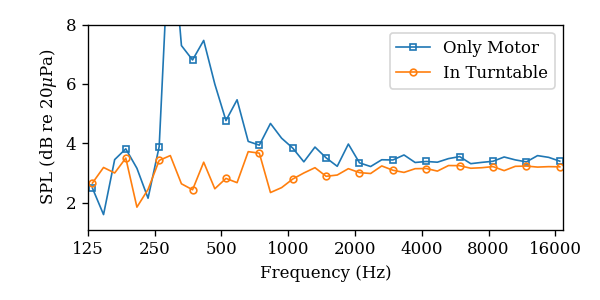}}
    \caption{ The 3D printed turntable damps the motor vibrations at 0.2 rev/s, reducing the radiated sound pressure. }
    \label{fig:turntable}
\end{figure}

Alongside the dummy head, the proposed turntable, shown in \cref{fig:subjects}, is also 3D printed. This increases the accessibility and cost efficiency of the overall device. As illustrated in \cref{fig:turntable}, the turntable structure does not significantly amplify the near-silent motor noise.

\section{Robot-enabled Experiments}
\label{sec:method}

The SNR gain is a fundamental objective performance metric for audio processing tasks such as speech enhancement or source separation. This metric indicates how well an audio algorithm performs by comparing the SNR of the output, processed audio with the input, raw audio. Larger SNR gain indicates better performance.
To calculate input/output SNR, engineers need to know the power of the target signal and noise/interference. With only the mixture audio, as in experiments with human actors, researchers cannot accurately calculate SNR. For this reason, researchers will often use loudspeakers in place of actors, as these are unmoving and repeatable sound sources.
For instance, one can arrange loudspeakers and microphones in a setting of interest. A first recording captures noise/interference and a second pass records the target loudspeakers. The recordings are scaled and summed to produce an artificial mix. If no equipment is moved between recordings, the mixed audio will accurately model a natural mixture of all the sources. A drawback is that ambient noise will be unrealistically amplified.

\begin{figure}[ht]
    \centering
    \centerline{\includegraphics[width=0.7\columnwidth]{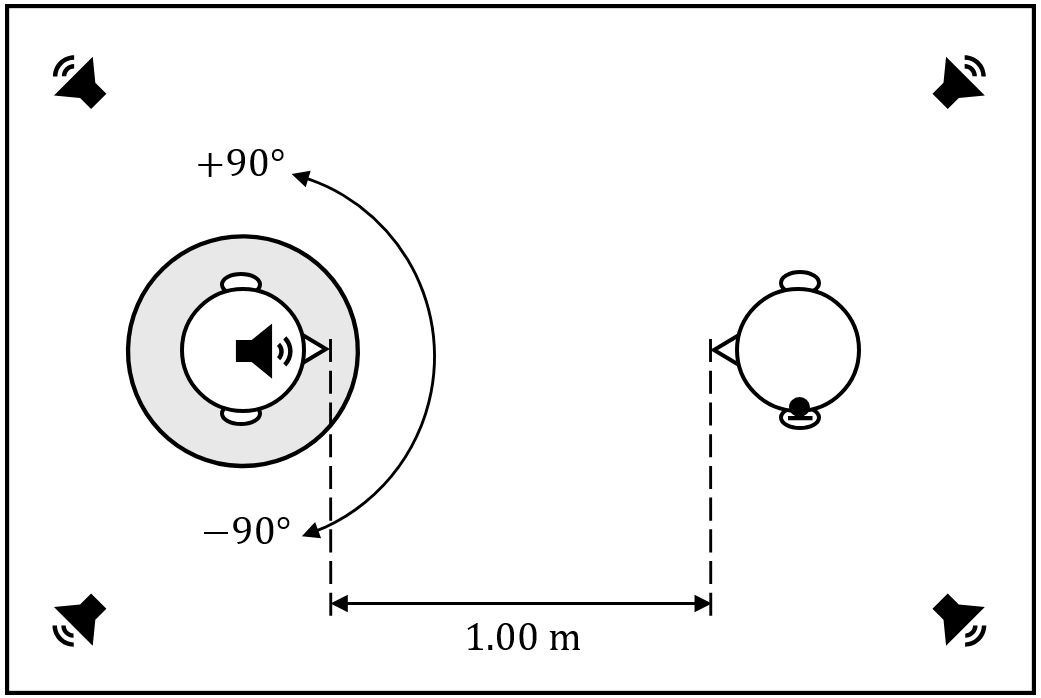}}
    \caption{ In the beamforming experiment, one printed head on a turntable simulated a moving talker while a stationary head simulated a listener. Both ear microphones are used, the left ear (shown) is targeted. Loudspeakers generated diffuse background noise. }
    \label{fig:setup}
\end{figure}

An experiment of this manner is performed with the robotic dummy head in an acoustically treated recording space. Four loudspeakers, facing the corners of the room, produce diffuse-like noise as in \cite{fejgin2021comparison}, and a 3D-printed dummy head emits target ``speech'' sounds. White noise is used as the source and noise signals, and re-used between recordings. A second dummy head ``listens'' through two ear microphones as described previously. A diagram in \cref{fig:setup} shows the experimental setup. All sound sources are placed at approximately equal height.

\begin{figure}[ht]
    \centering
    \centerline{\includegraphics[trim={0 0.3cm 0 0},clip,width=0.72\columnwidth]{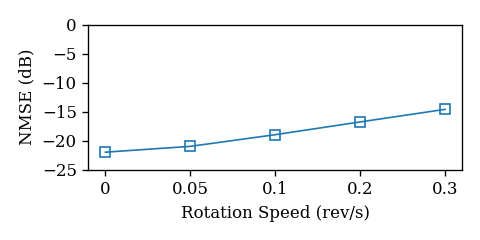}}
    \caption{ The sample-wise normalized mean-squared error (NMSE) between the artificially- and physically-mixed waveforms shows a high level of agreement at all source speeds. At 0 rev/s, the NMSE is attributed to the 23.2 dB SNR ambient noise. Recordings are taken at a sampling frequency of 48 kHz.}
    \label{fig:mixing_acc}
\end{figure}

\begin{figure}[ht]
    \centering
    \centerline{\includegraphics[trim={0 0.3cm 0 0},clip,width=0.7\columnwidth]{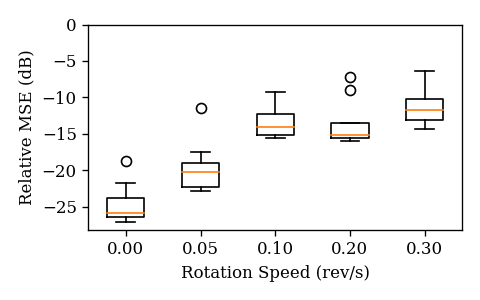}}
    \caption{
    In the beamforming experiment, one printed head on a turntable simulated a moving talker while a stationary head simulated a listener. Both ear microphones were used. Only the target microphone in the left ear is shown. Loudspeakers generated diffuse background noise.
    }
    \label{fig:rep}
\end{figure}

Two types of scenario are studied: in the spatially-stationary case, the talker is unmoving, facing the listener as shown in \cref{fig:setup}; in the dynamic cases, the talker starts facing $-90^\circ$ then makes one rotation to $+90^\circ$ at a constant speed. Various reasonable head rotation speeds are considered.
For each scenario, three recordings are acquired: a noise-only recording, target speech, and a natural mixture, in no particular order. In \cref{fig:mixing_acc}, it is shown that artificially mixing the former two signals closely mimics the latter, with the added benefit that various other SNR conditions can be simulated without performing extra recordings. While there appears to be some error proportional to the rate of motion, the overall error is comparable to error from uncontrolled ambient noise.
To evaluate repeatability, the target speech recordings are repeated 8 times for each target source speed. Sample-wise relative mean-squared error is calculated using the sample-wise average recording as reference to reveal a high degree of repeatability in \cref{fig:rep}.

These benchmark results in \cref{fig:mixing_acc} and \cref{fig:rep} confirm that the spatially-dynamic, robot-enabled recordings are repeatable and therefore suited for the objective evaluation of audio algorithms.

\section{Application to Binaural Beamforming}
\label{sec:experiment}

To illustrate how these repeatable dynamic recordings can be used in audio signal processing research, the proposed device is used to evaluate a motion-robust beamformer based on \cite{gossling2019rtf} with a moving target talker and unmoving dfTeX error: pdflatex (file ./png/ilds.png): xpdf: reading PDF image failedlistener. This section uses the artificially mixed audio of Section \ref{sec:method} scaled to 10 dB SNR.

\subsection{Signal model}
The audio received by the $M=2$ microphones in the listener's ears is modeled as a complex vector in the STFT domain \cite{gannot2017consolidated}
\begin{align}
\label{eq:atf}
    \mv{x}[k,l] &= \mv{a}[k,l] s[k,l] + \mv{v}[k,l],
\end{align}
\noindent
where $k,l$ are indices for time frames and frequency bins respectively, $\mv{a}[k,l]$ is the acoustic transfer function, $s[k,l]$ is the STFT of the target talker's ``speech'', and $\mv{v}[k,l]$ is spatially uncorrelated noise. Note that the acoustic transfer function is time-varying to account for a dynamic scenario, i.e. motion. The STFT is taken with a frame length of 20 ms, 50\% overlap, and a root-Hann window.

The target signal is the reverberated, perceptually-informative source image $d[k,l] = a_0[k,l] s[k,l]$, i.e. the target talker's speech as received by microphone $m=0$, which is assigned without loss of generality as the left ear microphone. Repeating this processing for the right ear would produce spatialized audio. Equation \eqref{eq:atf} is rewritten as
\begin{align}
\label{eqn:signal}
    \mv{x}[k,l] &= \mv{h}[k,l] d[k,l] + \mv{v}[k,l],
\end{align}

\noindent
where $\mv{h}[k,l] = \mv{a}[k,l]/a_0[k,l]$ is the relative transfer function (RTF).

\subsection{Denoising beamformer}

The Minimum Variance Distortionless Response (MVDR) beamformer has analytic solution \cite{vincent2018audio}
\begin{align}
    \mv{w}[k,l] = \frac{\mmat{R}_\mv{v}^{-1}[k,l] \mv{h}[k,l]}{\mv{h}\herm[k,l] \mmat{R}_\mv{v}^{-1}[k,l] \mv{h}[k,l]},
\end{align}
\noindent
where $\mmat{R}_\mv{v}[k,l] = E[\mv{v}[k,l] \mv{v}\herm[k,l]]$ is the noise covariance matrix. Using this filter in practice requires estimates of $\mmat{R}_\mv{v}[k,l] = E[\mv{v}[k,l] \mv{v}\herm[k,l]]$ and $\mv{h}[k,l]$.
As this experiment deals in motion, and the noise is stationary in time and space, an offline trained estimate $\widehat{\mmat{R}}_\mv{v}$ is used so only the spatial parameter $\widehat{\mv{h}}$ is adapted.

Per the covariance whitening (CW) method for RTF estimation, the input signal is first whitened\cite{markovich2015performance}
\begin{equation}
    \label{eq:whiten}
    \mv{y}[k,l] = 
    \widehat{\mmat{R}}_\mv{v}^{-1/2}[k,l]
    \mv{x}[k,l],
\end{equation}
\noindent
where $\widehat{\mmat{R}}_\mv{v}^{1/2}[k,l]$ is calculated by Cholesky decomposition. The whitened spatial covariance matrix (SCM) is estimated as
\begin{align}
    \label{eq:recursion}
    \widehat{\mmat{R}}_\mv{y}[k,l] = \alpha \widehat{\mmat{R}}_\mv{y}[k-1,l] + (1-\alpha) \mv{y}[k,l] \mv{y}[k,l]\herm,
\end{align}
\noindent
where $\alpha$ is the forgetting factor corresponding to time-constant $\tau\in(0,1]$ s, as in \cite{gossling2019rtf}. In this work $\tau=200$ ms.
The RTF estimate is
\begin{equation}
    \label{eq:cw}
    \widehat{\mv{h}}=
    \frac{
    \widehat{\mmat{R}}_\mv{v}^{1/2}[k,l]
    \widehat{\mv{q}}[k,l]
    }{
    \mv{e}_1 \herm 
    \widehat{\mmat{R}}_\mv{v}^{1/2}[k,l]
    \widehat{\mv{q}}[k,l]
    },
\end{equation}
\noindent
where $\mv{e}_1\herm=\begin{bmatrix} 1\; 0 \;\cdots \, 0 \end{bmatrix}$ and $\widehat{\mv{q}}[k,l]$ is the principle eigenvector of $\widehat{\mmat{R}}_\mv{y}[k,l]$. Diagonal loading is not used as the SCM is well-conditioned.

\subsection{Objective performance vs. speed}

\begin{figure}[ht]
    \centering
    \centerline{\includegraphics[trim={0 0.3cm 0 0},clip,width=0.88\columnwidth]{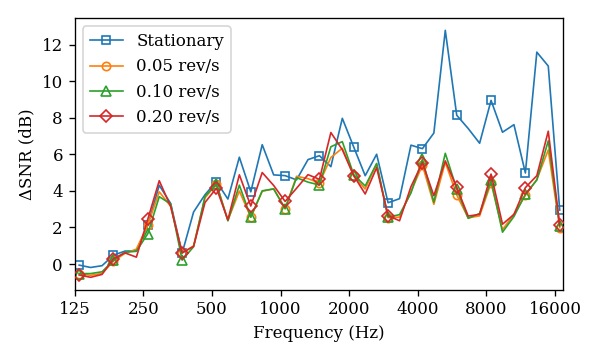}}
    \caption{ The SNR gain across frequency of the MVDR+CW beamformer, applied to binaural audio from a listener targeting a still/moving talking head. 
    The presence of any motion causes a severe drop in performance at high frequencies, while the rate of motion has a comparatively insignificant effect. }
    \label{fig:mvdr}
\end{figure}

Applying the MVDR+CW beamformer to audio recorded for various talker speeds reveals a surprising result. Beamforming performance  appears to vary depending on the \textit{presence} -- rather than the rate -- of motion, as illustrated in \cref{fig:mvdr}. We present this preliminary result as an example of the kind of interesting research that our device enables, and leave deeper analysis of this observation to future work.

\section{Conclusion}
\label{sec:conc}
In this work, we developed and deployed a robotic dummy head that is capable of simultaneous motion while recording. Additional benefits of this research tool are its low-cost and open-sourced design, which can be fabricated on standard, commonly-available 3D printers.

The robotic dummy head is relevant to various audio tasks, including but not limited to sound source localization and tracking, head pose estimation, source separation, and the cocktail party problem. 
More generally, the proposed device might be used to autonomously generate large-scale labeled datasets of spatial audio.
Overall, we anticipate that many exciting developments will be enabled by the robotic acoustic head simulator.

\bibliographystyle{IEEEbib}
\bibliography{main}

\end{document}